\documentclass[pra,twocolumn,showpacs,preprintnumbers,amsmath,amssymb,superscriptaddress]{revtex4}

\usepackage{bm}
\usepackage{graphicx}
\usepackage{amsbsy}
\usepackage{amsmath}
\usepackage{amsfonts}
\usepackage{amsthm}

\begin{document}

\theoremstyle{plain}
\newtheorem{theorem}{Theorem}
\newtheorem{lemma}[theorem]{Lemma}
\newtheorem{corollary}[theorem]{Corollary}
\newtheorem{conjecture}[theorem]{Conjecture}
\newtheorem{proposition}[theorem]{Proposition}

\theoremstyle{definition}
\newtheorem{definition}{Definition}

\theoremstyle{remark}
\newtheorem*{remark}{Remark}
\newtheorem{example}{Example}

\title{Mixed State Entanglement of Assistance and the Generalized Concurrence}     
\author{Gilad Gour}\email{ggour@math.ucsd.edu}
\affiliation{Department of Mathematics, University of California/San Diego, 
        La Jolla, California 92093-0112}

\date{\today}

\begin{abstract} 
We consider the maximum bipartite entanglement that can be distilled
from a single copy of a multipartite mixed entangled state, where we focus 
mostly on $d\times d\times n$-dimensional tripartite mixed states. We show 
that this {\em assisted entanglement}, when measured in terms of the generalized 
concurrence (named G-concurrence) is (tightly) bounded by an entanglement monotone, which 
we call the G-concurrence of assistance. The G-concurrence is one of the possible 
generalizations of the concurrence to higher dimensions, and for pure bipartite states it 
measures the {\em geometric mean} of the Schmidt numbers. For a large (non-trivial) class 
of $d\times d$-dimensional mixed states, we are able to generalize Wootters
formula for the concurrence into lower and upper bounds on the G-concurrence.  
Moreover, we have found an explicit formula for the G-concurrence of assistance
that generalizes the expression for the concurrence of assistance
for a large class of $d\times d\times n$-dimensional tripartite pure states.  
\end{abstract} 

\pacs{03.67.Mn, 03.67.Hk, 03.65.Ud}

\maketitle

\section{Introduction}

Preparation of entanglement between distant parties is
an important task required for quantum communication and 
quantum information processing~\cite{NC}. In the last years there has been
an intensive research in the field of quantum communication which yields
a variety of methods to distribute bipartite entanglement, such as 
entanglement swapping~\cite{Zuk93}, quantum repeaters~\cite{Bri98},
entanglement of assistance~\cite{DiV98,Coh98}, localizable entanglement~\cite{Ver04} and 
remote bipartite entangled state preparation~\cite{Gou04,Gou05}.  
Nevertheless, due to the lack of a complete
understanding of mixed state entanglement and multi-partite 
entanglement, it is not always clear what is the optimal way
to distribute entanglement among distant parties.

For example, suppose a {\em supplier} of entanglement, named ``Sapna'',
wishes to create entanglement between two distant parties, say, Alice
and Bob. Suppose also that Sapna creates the entanglement between Alice 
and Bob in two steps: first, she starts with three entangled qudits
in a pure state (such as the GHZ state) and sends, via quantum channels, 
two of the three qudits to Alice and Bob. Due to the operations of the 
quantum channels, the pure tripartite entangled state degrades to a mixed tripartite
entangled state. Second, she performs a measurement on her qudit and sends
the classical result to Alice and Bob. In this way, she can {\em assist}
Alice and Bob to increase their shared entanglement. 

For the more simple case in which Alice, Bob and Sapna sharing a {\em pure} tripartite 
state, the maximum average of entanglement that Sapna can prepare between Alice and Bob 
is called ``Entanglement of Assistance'' (EoA)~\cite{DiV98,Coh98}.  
Very recently~\cite{Win05,Hor05}, 
an explicit formula for the EoA has been found in the asymptotic 
limit of many copies of tripartite pure states. However, for a single copy, 
in general, 
it is very hard to find Sapna's optimal measurement.
Moreover, the 
optimal measurement depends on the choice of the measure of entanglement, and 
therefore it is extremely helpful to work with computationally manageable measure 
of entanglement.    

In this paper, we find a tight upper bound on the amount of bipartite entanglement
that can be distilled from a single copy of a $d \times d \times n$ 
tripartite mixed state. For $d=2$ our bound is given in terms of the convex roof extension
of the concurrence of assistance~\cite{Lau01}, and for $d>2$ in terms of an entanglement monotone which we
call the G-concurrence~\cite{Gou05} since for a pure bipartite state it is equal to the 
{\em geometric mean} of the Schmidt numbers. We show that the G-concurrence (GC) is a 
computationally manageable measure of entanglement, and in fact, for a large class of 
states, we find an {\em explicit} formula for the G-concurrence of assistance (GCoA), as well 
as lower and upper bounds for the GC. These formulas, naturally generalize the concurrence of 
assistance~\cite{Lau01} and Wootters 
formula~\cite{Woo98} to higher dimensions. Furthermore, the generalization of the GCoA to 
multipartite mixed states yields an entanglement monotone which provides an upper bound on the 
localizable entanglement of the parties. 

The paper is organized as follows: In section II we define assisted entanglement as the 
distillable bipartite entanglement from a single copy of tripartite mixed state and discuss its 
difference from entanglement of assistance. We then define the GCoA as well as the assisted 
G-concurrence. In section III, we show that the GCoA is an entanglement 
monotone and in section IV we show that it provides an upper bound on the assisted entanglement.  
Then, in section V we find explicit formulas and
bounds for the GC and the GCoA. Finally, in section VI we end with a short summary and conclusions.   

\section{Assisted entanglement verses entanglement of assistance}  

\subsection{Entanglement of Assistance of pure states}

For a pure tripartite state, $|\psi\rangle_{ABS}$, the EoA is defined by~\cite{DiV98,Coh98}:
\begin{equation}
	E_\text{a}(|\psi\rangle_\text{ABS})\equiv 
	\text{max}\sum_k p_k E(|\phi_k\rangle_\text{AB})\;, 
\label{eq:def}
\end{equation}
which is maximized over all possible decompositions of 
$\rho_\text{AB}=\sum_k p_k|\phi_k\rangle_\text{AB}\langle\phi_k|$, where
$\rho_\text{AB}\equiv\text{Tr}_{{}_{S}}\left(|\psi\rangle_\text{ABS}\langle\psi|\right)$.
In general, a distribution of states that maximizes Eq.~(\ref{eq:def})
for a given entanglement measure~$E$ will not necessarily be the optimal distribution
for a different measure. Therefore, the choice of measure is important
and depends on the planned quantum information task by Alice and Bob subsequent to
Sapna's assistance. 

Eq.~(\ref{eq:def}) provides the maximum average entanglement 
that Sapna can create between Alice and Bob because
any decomposition of $\rho_\text{AB}$ can be realized by a generalized measurement 
performed by Sapna~\cite{Hug93}. To see that 
(for more details see~\cite{Hug93}), let us first write $|\psi\rangle_{ABS}$ in the 
following Schmidt decomposition:
$
|\psi\rangle_{ABS}=\sum_{k=1}^{n}\sqrt{p_k}|\phi_k\rangle_{AB}|k\rangle_{S}\;,
$
where $n\leq d$ is the number of positive (non-zero) Schmidt numbers, $p_k$. Hence, Alice and Bob 
share the state
$
\rho_{AB}\equiv\sum_{k=1}^{n}p_k|\phi_k\rangle_{AB}\langle\phi_k|\;.
$
Now, for a given $m\times m$ unitary matrix $U$ ($m\geq n$), we define the following 
$m$ (possibly non-normalized and non-orthogonal) states: 
\begin{equation}
|v_l\rangle_{S}\equiv\sum_{k=1}^{n}U_{lk}|k\rangle_{S}\;,\;\;\;l=1,2,...,m\;.
\end{equation}
Since $U$ is unitary we have $\sum_{l=1}^{m}E_l=1$, with $E_l\equiv |v_l\rangle\langle v_l|$;
that is, the operators $\{E_l\}$ form a POVM with $m$ elements. 
We take this POVM to describe Sapna's measurement. It can be shown (see~\cite{Hug93}) 
that any decomposition of $\rho_{AB}$ can be realized by this measurement of Sapna with an
appropriate choice of $U$. Note that  
this POVM can be realized by a projective von-Neumann measurement on the original system {\em plus}
ancilla.  
 
\subsection{Assisted Entanglement}

Here we define {\em assisted entanglement}, $A_{E}$, as the maximum average of  
entanglement (measured with $E$) that Alice and Bob can share after a general {\em tripartite} LOCC.  
That is, in general, it is not known if the EoA, as defined in Eq.~(\ref{eq:def}), always provides 
the maximum average of entanglement under {\em general} three-party LOCC (i.e.\ not only
by Sapna's measurement). Thus, for a single copy of pure tripartite states, 
\begin{equation}
E_a \leq A_{E}\;.
\end{equation}
First, note that from its definition as the maximum distillable bipartite entanglement, $A_E$, 
is an entanglement monotone on tripartite pure {\em and} mixed 
states. 
Second, if there exist a measure, $E$, for which $E_a < A_E$ for some pure state, then in this case
the EoA when measured with $E$ can not be considered as an entanglement monotone for
tripartite pure states. Thus, it is very important to find measures of entanglement
such that for pure states $E_a=A_E$. After the next subsection we show that the GC satisfies this 
requirement.

\subsection{Definitions and notations}

For a pure bipartite state, $|\psi\rangle$, the GC is defined as 
the {\em geometric mean} of the (non-negative) Schmidt numbers
\begin{equation} 
G(|\psi\rangle)\equiv 
d(\lambda_{0}\lambda_{1}\cdots\lambda_{d-1})^{1\over d}
=d\left[{\rm Det}\left(A^{\dag}A\right)\right]^{1 \over d}\;,
\label{deta}
\end{equation}
where the matrix elements of $A$ are $a_{ij}$ 
($|\psi\rangle=\sum_{ij}a_{ij}|i\rangle|j\rangle$).
For a mixed $d\times d$-dimensional bipartite state, $\rho$, 
the GC is defined 
in terms of the convex roof extension:
\begin{equation}
G(\rho)={\rm min}\sum_{i}p_{i}G(|\psi _{i}\rangle)\;\;\left(\rho
=\sum_{i}p_{i}|\psi _{i}\rangle\langle\psi _{i}|\right)\;,
\label{CMM}
\end{equation}
where the minimum is taken over all decompositions of $\rho$. 
In~\cite{Gou05} it has been shown that the GC as defined in 
Eqs.~(\ref{deta},\ref{CMM}) is a bipartite entanglement monotone. 
We now show that it is possible to define a {\em tripartite} entanglement
monotone in terms of the GC.

\begin{definition}
(a) For pure $d\times d\times n$-dimensional tripartite states the G-concurrence of assistance
(GCoA), $G_a$, is defined by
\begin{equation}
G_a(|\psi\rangle_{ABS})={\rm max}\sum_{i} p_i G(|\phi_i \rangle_{AB})\;,
\label{ga}
\end{equation}
where the maximum is taken over all the decompositions of $\rho_{AB}\equiv 
{\rm Tr}_{_{S}}|\psi \rangle_{ABS}\langle\psi |=\sum p_i|\phi_i \rangle_{AB}\langle\phi_i|$.\\
(b) For tripartite mixed states $\rho_{ABS}=\sum_{j}q_j|\psi_j \rangle_{ABS}\langle\psi_j |$, 
the GCoA is defined in terms of the convex roof extension: 
\begin{equation}
G_a(\rho_{ABS})={\rm min}\sum_{j}q_jG_a(|\psi _j\rangle_{ABS})\;,
\label{gc}
\end{equation}
where the minimum is taken over all decompositions, $\{q_j,\;|\psi _j\rangle_{ABS}\}$,
of $\rho_{ABS}$.
\end{definition}
Note that with this definition, only for {\em pure} tripartite states 
the GCoA is equal to the maximum possible average of G-concurrence that 
Alice and Bob can share after Sapna's measurement. 
 
\begin{definition}
The assisted G-concurrence, $A_G$, is defined as the maximum average of  
G-concurrence that Alice and Bob can share after a general 
{\em tripartite} LOCC. That is,  
\begin{equation}
A_G(\rho_{ABS})={\rm max}\sum p_k G(\sigma_{k}^{AB})
\end{equation}
where the maximum is taken over all possible probability distributions 
of mixed states, $\{p_k,\;\sigma_{k}^{AB}\}$, shared between Alice and Bob
after a general tripartite LOCC.
\end{definition}
Note that from its definition the assisted G-concurrence can not increase by 
a general tripartite LOCC. 
That is, it is an entanglement monotone. 

\section{The G-concurrence of assistance}

In this section we prove that the GCoA is an entanglement monotone.
In particular, it follows that for pure states, $A_{G}=G_{a}$. 
That is, the optimal tripartite LOCC protocol for preparing the maximum 
possible G-concurrence between Alice and Bob
consists of only a single measurement by Sapna.  

\begin{theorem}
The G-concurrence of assistance is an entanglement monotone for tripartite
mixed states.
\end{theorem}

For for $2\times 2\times n$-dimensional {\em pure} states this theorem has been
proven in~\cite{GMS}.
\begin{proof}
In order to prove that the GCoA is an entanglement monotone we need to prove
two conditions~\cite{Vid98}:\\ 
(i) For any quantum operation 
$\varepsilon _k$ performed (locally) by Alice, Bob or Sapna,
\begin{equation}
G_a(\rho_{ABS})\geq \sum_{k}p_k G_{a}(\rho_k)\;,
\end{equation}
where $p_k\equiv {\rm Tr}\;\varepsilon_k(\rho_{ABS})$ and 
$\rho_k\equiv \varepsilon_k(\rho_{ABS})/p_k$.\\ 
(ii) $G_a(\rho_{ABS})$ is a convex function of $\rho_{ABS}$.

Condition (ii) is trivially satisfied due to the definition of 
the GCoA in terms of the convex roof 
extension (see Eq.~(\ref{gc})).  
In order to prove condition~(i) we first assume that 
$\varepsilon _k$ is performed locally by Alice and that the initial 
state is pure; i.e. $\rho_{ABS}=|\psi\rangle _{ABS}\langle\psi|$. The most 
general local operation $\varepsilon _k$ is given in terms of the 
Kraus operators $\hat{M}_{k,j}$:
\begin{equation}
\varepsilon _k (\rho_{ABS})=\sum_{j}\hat{M}_{k,j}|\psi\rangle _{ABS}\langle\psi|\hat{M}_{k,j}^{\dag}\;,
\end{equation}
where $\sum_{k,j}\hat{M}_{k,j}^{\dag}\hat{M}_{k,j}\leq I_{A}$ and $I_A$ is the identity operator
in Alice system. 
As the GC of any bipartite state $|\phi\rangle_\text{AB}$
satisfies~\cite{Gou05} 
$G(\hat{A}\otimes\hat{B}|\phi\rangle)=|\text{Det}\hat{A}|^{2/d}
|\text{Det}\hat{B}|^{2/d}\;G(|\phi\rangle)$, we obtain
\begin{equation}
\sum_{k}p_kG_a(\rho_k) =\sum_{k,j}|\text{Det}\hat{M}_{k,j}|^{2/d}
G_a(|\psi\rangle _{ABS})\;.
\end{equation}
Thus, condition~(i) follows from
the geometric-arithmetic inequality:
\begin{align} 
\sum_{k,j}\left|\text{Det}(\hat{M}_{k,j})\right|^{2/d}
& =\sum_{k,j}\sqrt[d]{\text{Det}(\hat{M}_{k,j}^{\dag}\hat{M}_{k,j})}\nonumber\\
& \leq \tfrac{1}{d}\sum _{k,j}\text{Tr}\hat{M}_{k,j}^{\dag}\hat{M}_{k,j}\leq 1\;.
\end{align}

We now consider the case in which the local operation $\varepsilon _{k}$ 
is performed by Sapna.
We define 
\begin{equation}
|\phi _{jk}\rangle\equiv \frac{1}{\sqrt{q_{jk}p_{k}}}\hat{M}_{k,j}|\psi\rangle _{ABS}\;,
\end{equation}
where the normalization factor $q_{jk}$ is taken such that 
$\langle\phi _{jk}|\phi _{jk}\rangle=1$. Note that 
\begin{equation}
\rho_k =\sum_j q_{jk}|\phi _{jk}\rangle\langle\phi _{jk}|\;.
\end{equation}
Thus,
\begin{align}
& \sum_{k}p_kG_a(\rho_k)  \leq \sum_{k,j}p_k q_{jk} G_a(|\phi_{jk}\rangle)\nonumber\\
& =\sum_{j,k}G_a(\hat{M}_{k,j}|\psi\rangle _{ABS})\leq G_a(|\psi\rangle _{ABS}).
\end{align} 
Now, for any quantity $E$ which is defined on mixed states in terms of the convex 
roof extension, and that satisfy 
$E(|\psi\rangle _{ABS})\geq \sum_{k}p_k E(\rho_k)$ for any tripartite pure state,
it follows that also $E(\rho_{ABS})\geq \sum_{k}p_k E(\rho_k)$ (see theorem 2 in~\cite{Vid98}).
Thus, the GCoA is an entanglement monotone.
\end{proof}

The definition of the GCoA can be extended to multipartite mixed states. For example,
suppose Alice, Bob and two ``suppliers'' $S_1$ and $S_2$ sharing four entangled
qudits in the state $|\psi\rangle_{ABS_1S_2}$. We then define the GCoA of the four qudit 
state as follows:
\begin{equation}
G_a(|\psi\rangle_{ABS_1S_2})={\rm max}\sum_{k}p_kG_a(|\psi^{k}\rangle_{ABS_1})\;,
\label{def:four}
\end{equation}
where the maximum is taken over all decompositions of 
$\rho_{ABS_1}=\sum_{k}p_k |\psi^{k}\rangle_{ABS_1}\langle\psi^{k}|$, with
$\rho_{ABS_1}\equiv {\rm Tr}_{S_2}|\psi\rangle_{ABS_1S_2}\langle\psi|$. 
Using Eq.~(\ref{def:four}) we can define the GCoA for four-partite mixed states in 
terms of the convex roof extension ({\it cf} Eq.~(\ref{gc})). Using the same lines of 
the proof above, it is possible to show that 
with this definition, the GCoA of the four-partite mixed states is also an entanglement 
monotone. Therefore, in this way we can define the GCoA for {\em any} multipartite mixed 
states and show that, indeed, it is an entanglement monotone. The definition of GCoA for 
multipartite states is useful especially in the context of localizable entanglement~\cite{Ver04}.    
 
\section{Upper bound on the assisted G-concurrence} 

The GCoA provides an upper bound on the amount of the average entanglement that 
Alice and Bob can share: 

\begin{theorem}
Let Alice, Bob and Sapna share a tripartite mixed state $\rho_{ABS}$. 
Then, 
\begin{equation}
A_{G}(\rho_{ABS})\leq G_{a}(\rho_{ABS})\;.
\label{bound}
\end{equation}
\end{theorem}

\begin{remark}
The theorem above can trivially be extended to include more than 3 parties. For example,
the GCoA as define in Eq.~(\ref{def:four}) for four parties (even if sharing a {\em mixed} 
entangled state) provides an upper bound on the localizable 
entanglement that Alice and Bob can share after local operations by $S_1$ and $S_2$~\cite{GG}. 
\end{remark}  

\begin{proof}
Let $\{q_j,\sigma_{ABS}^{j}\}$ be a probability distribution of tripartite mixed
states obtained from $\rho_{ABS}$ after a general tripartite LOCC. Let us also assume
that the LOCC protocol is optimal in the sense that
\begin{equation}
A_{G}(\rho_{ABS})=\sum_{j}q_j G(\sigma_{AB}^{j})\;,
\end{equation}
where $\sigma_{AB}^{j}={\rm Tr}_{{}_{S}}\sigma_{ABS}^{j}$.
Now, since the GCoA is an entanglement monotone, 
\begin{equation}
G_a(\rho_{{}_{ABS}}) \geq \sum_{j}q_j G_a(\sigma_{{}_{ABS}}^{j})
 =\sum_{ij}q_jp_{ij}G_a(|\phi ^{ij}_{{}_{ABS}}\rangle)\;, 
\end{equation}
where for each $j$, $\{p_{ij},\;|\phi ^{ij}\rangle_{ABS}\}$ is an optimal decomposition
of $\sigma_{ABS}^{j}$. Note also that $\sigma^{j}_{AB}=\sum_{ij}p_{ij}t_{{}_{AB}}^{ij}$, where
$t_{{}_{AB}}^{ij}\equiv {\rm Tr}_{{}_{S}}|\phi ^{ij}\rangle_{ABS}\langle\phi ^{ij}|$. Thus,
\begin{equation}
G_a(\rho_{ABS})\geq \sum_{ij}q_jp_{ij}G(t_{{}_{AB}}^{ij})\geq \sum_{j}q_j G(\sigma^{j}_{AB})\;.
\end{equation} 
That is, $G_a(\rho_{ABS})\geq A_{G}(\rho_{ABS})$
\end{proof}

The bound above on the assisted GC is tight and, in fact, in some cases
$G_a = A_{G}$. As mentioned in the beginning of section III, 
$G_a = A_{G}$ for pure tripartite states. Also, in some cases
in which Alice, Bob and Sapna sharing a
``slightly'' mixed $2\times 2\times 2$-dimensional state, 
$G_a = A_{G}$ up to a second order~\cite{Ban05}.
Moreover, the bound in Eq.~(\ref{bound}) is stronger than the upper bound given in~\cite{Gou05} for
mixed state entanglement swapping. 

In mixed state entanglement swapping, Sapna shares with Alice the state 
$\rho_{AS_1}$ and with Bob the state $\rho_{BS_2}$ (Sapna holds systems $S_1$ and $S_2$). 
In~\cite{Gou05} it has been shown that the maximum average GC, $A_G$, 
that Alice and Bob can share after any tripartite LOCC is bounded by 
$A_G \leq G(\rho_{AS_1})G(\rho_{BS_2})$, and equality is achievable if $\rho_{AS_1}$
and $\rho_{BS_2}$ are pure. We now prove a stronger version of this result.
\begin{proposition} 
\begin{equation}
G_{a}(\rho_{AS_1}\otimes\rho_{BS_2})\leq G(\rho_{AS_1})G(\rho_{BS_2})\;.
\end{equation}
\end{proposition}
Thus, in general, $G_a$ provides a tighter upper bound on $A_G$. 

\begin{proof}
Let $\{p_i,|\psi _i\rangle\}$ and $\{q_j,|\phi _j\rangle\}$ be
optimal decompositions of $\rho_{AS_1}$ and $\rho_{BS_2}$, respectively, such 
that $G(\rho_{AS_1})=\sum_{i} p_i G(\psi _i)$ and 
$G(\rho_{BS_2})=\sum_{j}q_j G(\phi _j)$. Thus,
\begin{align}
G_{a}(\rho_{AS_1}\otimes\rho_{BS_2}) & \leq 
\sum_{i,j}p_iq_jG_a(|\psi _i\rangle|\phi _j\rangle)\nonumber\\
&=G(\rho_{AS_1})G(\rho_{BS_2})\;,
\end{align}
where we have used the equality $G_a(|\psi _i\rangle|\phi _j\rangle)=G(\psi _i)G(\phi _j)$
(see~\cite{Gou05}).
\end{proof} 

\section{Explicit formulas and bounds}

In this section we find explicit formulas and bounds for the GC 
and the GCoA of a {\em pure} $d\times d\times n$-dimensional pure state
$|\psi\rangle_{ABS}$. We denote by $\rho$ the $d\times d$-dimensional mixed
state shared between Alice and Bob after tracing over Sapna's system.
For the GCoA of $\rho$ we find an explicit formula for a non-trivial (large) 
class of states which generalize to higher dimensions the formula for the 
concurrence of assistance~\cite{Lau01}. For the GC of $\rho$ we find lower and 
upper bounds that generalize to higher dimensions Wootters formula for the 
concurrence~\cite{Woo98}. 
 
We begin with a complete set of orthogonal $d\times d$-dimensional 
eigenstates, $|\phi_k\rangle$, corresponding to the non-zero eigenvalues
of $\rho$, and sub-normalized such that $\langle\phi_k|\phi_k\rangle$
is the $k$th eigenvalue of $\rho$. That is, 
\begin{equation}
\rho=\sum_{k=1}^{n}|\phi_k\rangle\langle\phi_k|\;,
\label{phi}
\end{equation}
where $n\leq d^{2}$ is the rank of $\rho$. Any other decomposition of
$\rho=\sum_{l=1}^{m}|\chi_l\rangle\langle\chi_l|$ is given by~\cite{Woo98}
\begin{equation}
|\chi_l\rangle=\sum_{k=1}^{n}U_{lk}^{*}|\phi_k\rangle\;,
\label{deco}
\end{equation}
where $m\geq n$ and $U$ is an $m\times m$ unitary matrix. We would like
to emphasize here that the action of the unitary group is not on density matrices 
but rather on different ensembles of the {\em same} density matrix $\rho$.

When $d=2$, one can define a symmetric ``tensor'' of rank 2 (see~\cite{Woo98}):
\begin{equation}
\tau ^{\phi}_{kk'}\equiv \langle\phi_k|\tilde{\phi}_{k'}\rangle\;,
\label{original}
\end{equation}
where $|\tilde{\phi}_{k'}\rangle\equiv\sigma_y
\otimes\sigma_y|\phi_{k'}^{*}\rangle$. We call $\tau ^{\phi}$ a 
tensor, because under a change of decomposition $\tau ^{\phi}$
transforms like a tensor. That is, for a general decomposition 
(see Eq.(\ref{deco})) we have (see also~\cite{Woo98})
\begin{equation}
\tau ^{{\chi}}_{ll'}\equiv \langle\chi_l|\tilde{\chi}_{l'}\rangle=
\sum_{k=1}^{n}\sum_{k'=1}^{n}U_{lk}U_{l'k'}\tau ^{\phi}_{kk'}\;.
\end{equation}
Moreover, since the tensor $\tau ^{\phi}$ is symmetric it can be 
diagonalized. We will show now that for $d>2$, similarly, one can define 
a completely symmetric tensor (i.e. invariant under index permutations)  
of rank $d$.

In order to generalize $\tau ^{\phi}_{kk'}$ (see Eq.~(\ref{original}))
into a tensor of rank $d$, let us first rewrite $\tau ^{\phi}_{kk'}$ in terms
of the coefficients $a_{ij}^{k}$, where
\begin{equation}
|\phi_{k}\rangle=\sum_{i=1}^{d}\sum_{j=1}^{d}a_{ij}^{k*}|i\rangle|j\rangle\;.
\label{phid}
\end{equation}
Then it is easy to check that Eq.~(\ref{original}) can be written as ($d=2$):
\begin{equation}
\tau ^{\phi}_{kk'}\equiv \langle\phi_k|\tilde{\phi}_{k'}\rangle
={\rm Det}
\begin{pmatrix}
a_{11}^{k} & a_{12}^{k} \cr
a_{21}^{k'} &  a_{22}^{k'} \cr
\end{pmatrix}
+{\rm Det}
\begin{pmatrix}
a_{11}^{k'} & a_{12}^{k'} \cr
a_{21}^{k} &  a_{22}^{k} \cr
\end{pmatrix}\;.
\label{original2}
\end{equation}
Thus, in analogy, for $d>2$, we define a completely symmetric tensor of rank $d$:
\begin{equation}
\tau ^{\phi}_{k_1k_2...k_d}\equiv \frac{d^{d/2}}{d!}\sum_{\sigma}{\rm Det}
\begin{pmatrix}
a_{11}^{k_{\sigma(1)}} & a_{12}^{k_{\sigma(1)}} & \cdot & \cdot & \cdot & a_{1d}^{k_{\sigma(1)}} \cr
a_{21}^{k_{\sigma(2)}} &  a_{22}^{k_{\sigma(2)}} & \cdot & \cdot & \cdot & a_{2d}^{k_{\sigma(2)}} \cr
\cdot & \cdot & \cdot & \cdot & \cdot & \cdot \cr
\cdot & \cdot & \cdot & \cdot & \cdot & \cdot \cr
\cdot & \cdot & \cdot & \cdot & \cdot & \cdot \cr
a_{d1}^{k_{\sigma(d)}} &  a_{d2}^{k_{\sigma(d)}} & \cdot & \cdot & \cdot & a_{dd}^{k_{\sigma(d)}} \cr
\end{pmatrix}\;,
\label{defe}
\end{equation}
where $\sigma$ varies over all permutations on $d$ symbols and the factor $d^{d/2}/d!$ has been chosen
such that $\left|\tau ^{\phi}_{kk...k}\right|^{2/d}$ is the GC of the sub-normalized state
$|\phi_k\rangle$. It follows from its definition that $\tau ^{\phi}_{k_1k_2...k_d}$ is symmetric under any 
permutation of its indexes.
In Appendix A we show that 
under a change of a decomposition, $\tau ^{\phi}$ (in Eq.~(\ref{defe}))
transforms like a tensor. That is, for a general decomposition 
(see Eq.(\ref{deco})) we prove
\begin{equation}
\tau ^{{\chi}}_{l_1l_2...l_d}=
\sum_{k_1=1}^{n}\sum_{k_2=1}^{n}\cdots\sum_{k_d=1}^{n}
U_{l_1k_1}U_{l_2k_2}...U_{l_dk_d}\tau ^{\phi}_{k_1k_2...k_d}\;.
\label{ndeco}
\end{equation}
Moreover, note that the $G$-concurrence of the ensemble of 
Eq.~(\ref{phi}) is conveniently expressed as:
\begin{equation}
\sum_{k=1}^{n} G(|\phi_k\rangle)
=\sum_{k=1}^{n}\left|\tau ^{\phi}_{kk...k}\right|^{2/d}\;. 
\end{equation} 

Unlike tensors of rank 2, higher-order tensors (even completely symmetric) 
cannot be reduced in general to a diagonal form by 
orthogonal or unitary transformations. We therefore define 
a class of $d\times d$-dimensional density matrices, $\mathcal{D}$, for which the 
$\tau$-tensors are diagonalizable. That is, $\rho\in \mathcal{D}$ if there exist a unitary matrix,
$U$, such that $\tau^{\chi}$ as in Eq.~(\ref{ndeco}) is a diagonal tensor.
For $d=2$ all density matrices belong to $\mathcal{D}$. However, for $d>2$ it is not clear
what is the size of the class $\mathcal{D}$ and, in fact, very little is known on the 
diagonalization of a $d^2\times d^2\times\cdots\times d^2$ matrix. Nevertheless,
note that even though $\tau^{\phi}$ is a tensor of rank $d$ 
(represented by a $d^2\times d^2\times\cdots\times d^2$ matrix), it is constructed
from a $d\times d$-dimensional density matrix $\rho$. That is, it depends on
less than $d^{4}$ parameters, whereas a general tensor of rank $d$
depends on $~d^{2d}$ parameters.
Thus, among all the completely symmetric
tensors of rank $d$, the $\tau$-tensors are a very small group. 
We hope to report in the future on how much smaller is the class of 
{\em diagonal} $\tau$-tensors relative to the group of {\em all} 
$\tau$-tensors.  
 
In the following we assume that $\tau^{\phi}_{k_1k_2...k_d}$ can be diagonalized
and therefore, without the loss of generality, we take it to be diagonal; i.e.
\begin{equation}
\tau^{\phi}_{k_1k_2...k_d}=\lambda_{k_{1}}\delta_{k_1k_2}\cdots\delta_{k_1k_d}\;,
\end{equation}
where, in general, $\lambda_k$ are complex numbers. However, we can take a diagonal 
unitary matrix $U$ (see Eq.(\ref{deco})) with appropriate phases in the diagonal, in order to obtain
a diagonal $\tau$-tensor with $|\lambda_k|$ on the diagonal. Thus, without the loss of 
generality we assume that $\lambda_k$ are real and non-negative.

\subsubsection{The G-concurrence of assistance}

\begin{theorem}
If $\rho\in \mathcal{D}$ then
\begin{equation}
G_a(\rho)=\sum_{k=1}^{n}\lambda_k^{2/d}\;.
\end{equation}
\end{theorem}

\begin{proof}
We need to show that there is no decomposition with higher average of GC. 
The average G-concurrence of a general decomposition (see Eqs.~(\ref{deco},\ref{ndeco}))
is given by:
\begin{equation}
\langle G\rangle_{\chi}=\sum_{l=1}^{m}\left|\tau ^{\chi}_{ll...l}\right|^{2/d}
=\sum_{l=1}^{m}\left|\sum_{k=1}^{n}\left(U_{lk}\right)^{d}\lambda_{k}\right|^{2/d}\;.
\label{impo}
\end{equation}
Since $\left|a+b\right|^{r}\leq |a|^r+|b|^r$, where
$0\leq r\leq 1$ and $a$, $b$ are complex numbers with absolute value less than 1,
\begin{equation}
\langle G\rangle_{\chi}\leq 
\sum_{l=1}^{m}\sum_{k=1}^{n}\left|U_{lk}\right|^{2}\lambda_{k}^{2/d}=\sum_{k=1}^{n}\lambda_{k}^{2/d}\;.
\end{equation}
\end{proof}

\begin{example}
Suppose Alice, Bob and Sapna sharing the state
\begin{equation}
|\psi\rangle_{ABS}=\sqrt{p_0}|000\rangle+\sqrt{p_1}|111\rangle+\sqrt{p_2}|222\rangle+\sqrt{p_3}|333\rangle\;.
\end{equation}
Thus, Alice and Bob sharing the mixed state
\begin{equation}
\rho=\sum_{k=0}^{3}|\phi_k\rangle\langle\phi_k|\;,
\label{e2}
\end{equation}
where $|\phi_k\rangle\equiv \sqrt{p_k}|kkk\rangle$. According to Eq.~(\ref{defe}), 
the super symmetric $\tau$-tensor that
corresponds to the decomposition in Eq.~(\ref{e2}) is given by:
\begin{equation}
\tau ^{\phi}_{1234}=\frac{2}{3}\sqrt{p_1p_2p_3p_4}\;,
\end{equation}
where for any other set of indexes $(k_1,k_2,k_3,k_4)$ which is {\em not} 
a permutation of $(1,2,3,4)$, 
$\tau ^{\phi}_{k_1k_2k_3k_4}=0$. According to Eq.~(\ref{ndeco}), for a different
decomposition, $\chi$, we have
\begin{equation}
\tau ^{{\chi}}_{l_1l_2l_3l_4}=\tau ^{\phi}_{1234}\sum_{\sigma}
U_{l_1\sigma(1)}U_{l_2\sigma(2)}U_{l_3\sigma(3)}U_{l_d\sigma(d)}\;,
\end{equation}
where $\sigma$ varies over all the 24 permutations of 4 symbols.
For the orthogonal matrix
\begin{equation}
U=\frac{1}{2}
\begin{pmatrix}
1 & 1 & 1 & 1 \cr
1 & 1 & -1 & -1 \cr
1 & -1 & 1 & -1 \cr
1 & -1 & -1 & 1 \cr
\end{pmatrix}\;, 
\end{equation}
we get a diagonal tensor $\tau^{\chi}$. The 4 diagonal elements of $\tau^{\chi}$ are all
equal to $\lambda_{k}=\sqrt{p_1p_2p_3p_4}$ ($k=1,2,3,4$). Hence, from theorem 4
\begin{equation}
G_a(|\psi\rangle_{ABS})=4\left(p_1p_2p_3p_4\right)^{1/4}\;.
\end{equation}
\end{example}

\subsubsection{The G-concurrence}

In order to find the decomposition with the least average of GC we first prove the following lemma:
\begin{lemma}  
If $\rho\in \mathcal{D}$, for any $d^2$ number of phases $\theta_1,\;\theta_2\;,...,\theta_{d^2}$, there exist
a decomposition such that the G-concurrence of each (sub-normalized) state in the decomposition 
is equal to $\left|\lambda_1\exp(i\theta_1)+\lambda_2\exp(i\theta_2)+...
+\lambda_{d^2}\exp(i\theta_{d^2})\right|^{2/d}/d^2$
(if $n<d^2$ then $\lambda_k\equiv 0$ for $n<k\leq d^2$).
\end{lemma}

\begin{proof} 
First, we take the unitary matrix $U_{lk}=\delta_{lk}\exp(i\theta_k/d)$ and 
substitute in Eq.(\ref{impo}) to get a decomposition with diagonal $\tau$-tensor whose diagonal elements are 
$\lambda_k\exp(i\theta _k)$. Then, we take a $d^2\times d^2$-dimensional unitary matrix $U$ with the property
that $\left(U_{lk}\right)^{d}=1/d^{d}$ for {\em all} $k$ and $l$. It is left to show that such a unitary 
matrix exists. In order to show that, we first define a set of $d^{2}$ $d\times d$-unitary 
matrices $V^{(a,b)}$ ($a,b=1,...,d$) with elements  
\begin{equation}
V^{(a,b)}_{jj'}={1\over d}\exp\left(\frac{2\pi i}{d}(j+a)(j'+b)\right)\;
; \;j,j'=1,2,...,d\;.
\end{equation}
Then we define a $d^2\times d^2$ matrix $U$:
\begin{equation}
U\equiv
\begin{pmatrix}
V^{(1,1)} & V^{(1,2)} & \cdot & \cdot & V^{(1,d)} \cr
V^{(2,1)} & V^{(2,2)} & \cdot & \cdot & V^{(2,d)} \cr
\cdot & \cdot & \; & \; & \cdot & \cr
\cdot & \cdot & \; & \; & \cdot & \cr
V^{(d,1)} & V^{(d,2)} & \cdot & \cdot & V^{(d,d)} \cr
\end{pmatrix}
\end{equation}
Note that each $V^{(a,b)}$ above is a $d\times d$ matrix. It is easy to check that
the matrix $U$ as defined above is a $d^2\times d^2$ unitary matrix with the desired property.
\end{proof}

In the following, without the loss of generality, we assume that $\lambda_{1}\geq \lambda _{k}$
for all $k$.

\begin{corollary} 
If $\rho\in \mathcal{D}$ and $\lambda _1\leq\lambda_2+...+\lambda_n$, then $G(\rho)=0$.
\end{corollary}

Note that the corollary above provides a necessary condition for separability
of bipartite mixed states.
 
\begin{theorem} 
If $\rho\in \mathcal{D}$ and $\lambda _1>\lambda_2+...+\lambda_n$ 
\begin{equation}
\lambda_{1}^{2/d}-\sum_{k=2}^{n}\lambda_{k}^{2/d} \leq
G(\rho)\leq\left|\lambda_1-\sum_{k=2}^{n}\lambda_k\right|^{2/d}\;.
\end{equation}
\end{theorem}
\begin{proof}
The upper bound follows
from the lemma above for $\theta_1=0$ and $ \theta_l=\pi$ ($2\leq l\leq n$). 
In order to prove the lower bound, first note that 
\begin{equation}
\left||a|^{r}-|b|^{r}\right|\leq\left|a+b\right|^{r}\leq |a|^r+|b|^r
\end{equation}
where $0\leq r\leq 1$ and $a$, $b$ are complex numbers with absolute value less than 1.
Hence,
\begin{align}
\sum_{l=1}^{m}\left|\sum_{k=1}^{n}\left(U_{lk}\right)^{d}\lambda_{k}\right|^{2 \over d}
& \geq \sum_{l=1}^{m}\left(|U_{l1}|^{2}\lambda_1^{2\over d}
-\left|\sum_{k=2}^{n}\left(U_{lk}\right)^{d}\lambda_{k}\right|^{2 \over d}\right)\nonumber\\
& \geq \lambda_{1}^{2/d}-\sum_{k=2}^{n}\lambda_{k}^{2/d}\;.
\end{align}
\end{proof}

Note that for $d=2$ both the lower and upper bounds reduce to Wootters formula of the 
concurrence~\cite{Woo98}. Also, if $\lambda_{k}=0$ for $k\geq 2$, then 
$G(\rho)=G_a(\rho)=\lambda_{1}^{2/d}$.
\begin{example} 
Consider the $3\times 3$-dimensional mixed state
\begin{equation}
\rho=p|\chi\rangle\langle\chi|+(1-p)|01\rangle\langle 01|\;,
\end{equation}
where $|\chi\rangle=(|00\rangle+|11\rangle+|22\rangle)/\sqrt{3}$ is a maximally
entangled state. It is easy to check that for this state $\lambda_{2}=\lambda_{3}=0$ 
and $\lambda_{1}=p^{3/2}$. Therefore, in this case $G=G_a=p$.
\end{example}

\section{Summary and Conclusions}

In summary, we have considered the maximum bipartite entanglement that can be distilled
from a single copy of a multipartite mixed entangled state, where we focused mostly on 
tripartite mixed states. We have shown that this 
{\em assisted entanglement}, when measured in terms of the G-concurrence 
(as defined in~Eqs.(\ref{deta},\ref{CMM})), is (tightly) bounded by the entanglement monotone 
given in Eqs.(\ref{ga},\ref{gc}), 
which we call the G-concurrence of assistance. The G-concurrence is one of the possible 
generalizations of the concurrence to higher dimensions, and for pure bipartite states it 
measures the {\em geometric mean} of the Schmidt numbers. For a large (non-trivial) class, 
$\mathcal{D}$, of $d\times d$-dimensional mixed states, we where able to generalize Wootters
formula for the concurrence, where the concurrence is replaced with the
G-concurrence. Moreover, we have found an explicit formula for the G-concurrence of assistance
for tripartite pure states, $|\psi\rangle_{ABS}$, with $\rho_{AB}\in \mathcal{D}$
($\rho_{AB}\equiv{\rm Tr}_{s}|\psi\rangle_{ABS}\langle\psi |$). 

In addition to GC and GCoA being valuable for finding upper bounds for the assisted 
entanglement of three (or more) parties,  
for $2\times2\times2$-dimensional pure states the concurrence
as well as the concurrence of assistance exhibits
monogamy constraints (entanglement tradeoffs)~\cite{Cof00,GMS}. 
Similarly, we are willing to conjecture that the generalizations
to a $d \times d \times d$ dimensional pure state, $|\psi\rangle_\text{ABS}$,
also holds:
\begin{conjecture}
\begin{align}
& \left[G(\rho_{AB})\right]^d+\left[G(\rho_{AS})\right]^d\leq \left[G_{A(BS)}\right]^d\nonumber\\	
& \left[G_{\text{a}}(\rho_{AB})\right]^d+\left[G_{\text{a}}(\rho_{AS})\right]^d\geq \left[G_{A(BS)}\right]^d .
\label{11}
\end{align}
\end{conjecture}

\emph{Acknowledgments:---}
I appreciate valuable discussions with S.~Bandyopadhyay, M.~S.~Byrd,
J.~Eisert, B.~C.~Sanders, R.~W.~Spekkens and N.~R.~Wallach. 
The author acknowledge support by the DARPA 
QuIST program under Contract No.\ F49620-02-C-0010 and the NSF 
under Grant No.\ ECS-0202087.

\appendix
\section{}

In this appendix we prove that the function, $\tau ^{\phi}_{k_1k_2...k_d}$, 
as define in Eq.~(\ref{defe}) is indeed a tensor.
That is, under a change of a decomposition, $\tau ^{\phi}$ transforms as in 
Eq.~(\ref{ndeco}). 

In order to prove Eq.~(\ref{ndeco}) we start with the definition of 
$\tau ^{\phi}$:
\begin{equation}
\tau ^{\phi}_{k_1k_2...k_d}=\frac{d^{d/2}}{d!}\sum_{\sigma}\sum_{\pi}{\varepsilon}_{\pi}
a_{1\pi(1)}^{k_{\sigma(1)}}a_{2\pi(2)}^{k_{\sigma(2)}}\cdots 
a_{d\pi(d)}^{k_{\sigma(d)}}\;,
\end{equation}
where $\pi$ and $\sigma$ varies over all permutations on $d$ symbols and
$\varepsilon_{\pi}$ is the signature of the permutation $\pi$. 

Now, from Eqs.(\ref{deco},\ref{phid}) it follows that:
\begin{equation}
|\chi_{l}\rangle=\sum_{i=1}^{d}\sum_{j=1}^{d}b_{ij}^{l}|i\rangle|j\rangle
\;,\;\;b_{ij}^{l}\equiv\sum_{k=1}^{n}U_{lk}a_{ij}^{k}\;.
\end{equation}
Thus, the tensor $\tau^{\chi}$ is given by:
\begin{align}
& \tau ^{\chi}_{l_1l_2...l_d} 
=\frac{d^{d/2}}{d!}\sum_{\sigma}\sum_{\pi}{\varepsilon}_{\pi}
b_{1\pi(1)}^{l_{\sigma(1)}}b_{2\pi(2)}^{l_{\sigma(2)}}\cdots 
b_{d\pi(d)}^{l_{\sigma(d)}}\nonumber\\
& = \frac{d^{d/2}}{d!}\sum_{\sigma}\sum_{\pi}{\varepsilon}_{\pi}
\sum_{k_1=1}^{n}\cdots\sum_{k_d=1}^{n}U_{l_{\sigma(1)}k_1}a_{1\pi(1)}^{k_1}\nonumber\\
& \;\;\;\;\;\;\;\;\;\;\;\;\;\;\;\;\;\;\;\;\;\;\;
\times U_{l_{\sigma(2)}k_2}a_{2\pi(2)}^{k_2}...U_{l_{\sigma(d)}k_d}a_{d\pi(d)}^{k_d}\;. 
\label{long}
\end{align}
Now, since the indexes $k_1,\;k_2,\;...,k_d$ are dummy indexes, for each permutation $\sigma$ 
in the above expression we can replace
the term 
$$
U_{l_{\sigma(1)}k_1}a_{1\pi(1)}^{k_1}
U_{l_{\sigma(2)}k_2}a_{2\pi(2)}^{k_2}\cdots 
U_{l_{\sigma(d)}k_d}a_{d\pi(d)}^{k_d}
$$ 
with 
$$
U_{l_{\sigma(1)}k_{\sigma(1)}}a_{1\pi(1)}^{k_{\sigma(1)}}
U_{l_{\sigma(2)}k_{\sigma(2)}}a_{2\pi(2)}^{k_{\sigma(2)}}\cdots 
U_{l_{\sigma(d)}k_{\sigma(d)}}a_{d\pi(d)}^{k_{\sigma(d)}}\;.
$$ 
However, since $\sigma$ is a permutation
we have 
\begin{equation}
U_{l_{\sigma(1)}k_{\sigma(1)}}U_{l_{\sigma(2)}k_{\sigma(2)}}\cdots U_{l_{\sigma(d)}k_{\sigma(d)}}
=U_{l_{1}k_{1}}U_{l_{2}k_{2}}\cdots U_{l_{d}k_{d}}\;.
\end{equation}
Thus, substituting all this in Eq.~(\ref{long}) gives
\begin{align}
& \tau ^{{\chi}}_{l_1l_2...l_d}=\frac{d^{d/2}}{d!}
\sum_{k_1=1}^{n}\sum_{k_2=1}^{n}\cdots\sum_{k_d=1}^{n}\sum_{\sigma}\sum_{\pi}\nonumber\\
& U_{l_1k_1}U_{l_2k_2}...U_{l_dk_d}{\varepsilon}_{\pi}
a_{1\pi(1)}^{k_{\sigma(1)}}a_{2\pi(2)}^{k_{\sigma(2)}}\cdots 
a_{d\pi(d)}^{k_{\sigma(d)}}\nonumber\\
& =\sum_{k_1=1}^{n}\sum_{k_2=1}^{n}\cdots\sum_{k_d=1}^{n}
U_{l_1k_1}U_{l_2k_2}...U_{l_dk_d}\tau ^{\phi}_{k_1k_2...k_d}\;.
\end{align}
That is, $\tau^{\phi}$ transforms like a tensor.


\begin{references}

\bibitem{NC}
M.~A.~Nielsen and I.~L.~Chuang, {\it ``Quantum Computation and Quantum Information''}
(Cambridge University Press, 2000).

\bibitem{Zuk93}
M.~\.{Z}ukowski, A.~Zeilinger, M.~A.~Horne, and A.~K.~Ekert,
\prl \textbf{71}, 4287 (1993);
S. Bose, V. Vedral, and P. L. Knight, \pra \textbf{57}, 822 (1998); 
\textbf{60}, 194 (1999).
B.-S. Shi, Y.-K. Jiang, G.-C. Guo,  \pra \textbf{62}, 054301 (2000);
L.~Hardy and D.~D.~Song, \pra \textbf{62}, 052315 (2000).

\bibitem{Bri98}
W.~D\"ur, H.-~J.~Briegel, J.~I.~Cirac and P.~Zoller, \pra \textbf{59}, 169 (1999).

\bibitem {DiV98} 
D.~P. DiVincenzo et al, 
	``The entanglement of assistance'', in Lecture Notes in Computer Science \textbf{1509}
	(Springer-Verlag, Berlin, 1999), pp. 247-257.

\bibitem {Coh98} O. Cohen, \prl \textbf{80}, 2493 (1998).

\bibitem{Ver04} 
F. Verstraete, M. Popp, and J. I. Cirac,
 \prl \textbf{92}, 027901 (2004); 
M. Popp, F. Verstraete, M. A. Martin-Delgado and J. I. Cirac
\pra \textbf{71}, 042306 (2005).

\bibitem{Gou04}
G.~Gour and B.~C.~Senders, \prl \textbf{93}, 260501 (2004).

\bibitem{Gou05}
G.~Gour, \pra, {\bf 71}, 012318 (2005).

\bibitem {Win05} J. A. Smolin, F. Verstraete and A. Winter, quant-ph/0505038.

\bibitem {Hor05} M. Horodecki, J. Oppenheim and A. Winter, quant-ph/0505062.

\bibitem {Lau01} T. Laustsen, F. Verstraete, and S. J. van Enk,
	Quant. Inf. and Comp. \textbf{3}, 64 (2003).

\bibitem{Woo98}
W.~K.~Wootters, \prl \textbf{80}, 2245 (1998).

\bibitem {Hug93} 
L. P. Hughston, R. Jozsa and W. K. Wootters, Phys. Lett. A \textbf{183}, 14 (1993).

\bibitem {Vid98} G. Vidal, J. Mod. Opt. \textbf{47}, 355 (2000).

\bibitem {GMS}
G.~Gour, D.~A.~Meyer and B.~C.~Sanders, quant-ph/0505091. 

\bibitem {GG}
G.~Gour, work in progress.

\bibitem {Ban05}
S.~Bandyopadhyay, G.~Gour, S.~Jami and B.~C.~Sanders, in preperation.

\bibitem {Cof00} 
V. Coffman, J. Kundu, and W. K. Wootters, \pra \textbf{61}, 052306 (2000).

\end{references}
\end{document}